\documentstyle[epsf]{elsart} %
\begin{document}
\begin{frontmatter}
\title{Localization Effect in a 2D Superconducting Network without 
Disorder} \author{B. Pannetier, C.C. 
Abilio, E. Serret, Th. Fournier and P. Butaud} 
\address{CNRS-CRTBT, associ\'e \`a l'Universit\'e Joseph Fourier\\
25 Av.  des Martyrs, 38042 Grenoble Cedex 9, France} \author{J. Vidal} 
\address{Groupe de Physique des Solides-CNRS, UMR 7588, 
Universit\'{e}s Paris 7 et Paris 6,\\
2 place Jussieu, 75251 Paris Cedex 05, France}
\begin{abstract}
The superconducting properties of a two-dimensional superconducting
wire network with a new geometry have been measured as a function of
the external magnetic field.  The extreme localization effect recently
predicted for this periodic lattice is revealed as a suppression of
the critical current when the applied magnetic field corresponds to
half a flux quantum per unit cell.  For this particular magnetic field, 
the observed vortex state configuration is highly disordered.
 
\end{abstract}
\end{frontmatter}
%
PACS[72.15, 73.23, 74.25]

\section{Introduction}
Superconducting wire networks are well known model systems which
provide an unique experimental access \cite{exp} to the fascinating
properties of the energy spectrum of tight binding electrons.  For
example the critical temperature of a superconducting network exhibits
an oscillatory dependence with the magnetic field which very
accurately describes the features of the ground state of the spectrum. 
The mapping of the superconducting transition line to the Landau
Levels is a direct consequence of the analogy between the linear
Ginzburg-Landau equation for the superconducting order parameter and
the Schroedinger equation for noninteracting charged particles.  Other
superconducting properties such as the equilibrium magnetization
\cite{Gandit} or the critical current \cite{Buisson} have also been
shown to result directly from the Landau levels spectrum.\\
 
Recently a novel case of extreme localization induced by a magnetic field was predicted \cite{Vidal} for
noninteracting electrons in a regular two dimensional lattice with the
so-called T3 geometry (Fig.1, inset).  This phenomenon is due to a
subtle interplay between the lattice geometry and the magnetic field
and occurs for a particular magnetic flux which corresponds to half a
flux quantum per plaquette.  For this special magnetic flux the
hopping terms between neighbour sites interfere destructively
and lead to a confinement of the electron motion within the so-called
Aharonov-Bohm cages.  The properties of these electronic states are
associated with an interesting behaviour of the Landau level spectrum
for the $T3$ lattice at half frustration $f=1/2$.  Here the
frustration $f=\phi/\phi_{0}$ is defined as the magnetic flux per plaquette in units of
the flux quantum $\phi_{0}=h/e$.  Instead of forming energy bands, the highly 
degenerate eigenvalues
merge into only 3 \underline {discrete} levels.  The confinement effect is a
consequence of the non-dispersive character of the eigenstates.  This
contrasts with the case of a square lattice \cite{Hofstadter} where
the eigenstates are dispersive and form broad energy bands
$\epsilon(k,f)$ at every rational frustration.\\
The signature of the finite group velocity $v=\frac {1}{\hbar}\frac
{\partial{\epsilon}}{\partial{k}}$ of such extended states for a
superconducting system manifests itself in the ability of the
superconducting wavefuntion to carry a supercurrent.  A simple model
based on the depairing current of a superconducting wire
\cite{Tinkham} was developped in \cite {Wang,Buisson} for
superconducting wire networks.  According to this model the critical
current exhibits strong maxima at rational frustrations $f=p/q$ (p and
q integer numbers) with strength proportional to the curvature of
$\epsilon (k)$.  The validity of this picture was confirmed by
experiments in a square lattice \cite{Buisson}.  An alternate
formulation in terms of vortex pinning for commensurate vortex lattice
at rational $f$ is fully equivalent.\\
The absence of dispersion $\epsilon (k)$ in the $T3$ lattice suggests
that the corresponding superconducting network should be unable to
carry a supercurrent at $f=1/2$.

\section{Experimental Results}
We have fabricated superconducting wire networks that reproduce the T3
geometry and we have studied the consequences of the above charge
confinement phenomenon on the superconducting properties : critical
temperature, critical current and vortex configuration.

\begin{figure}
	\epsfxsize=100mm
		  \centerline{\epsffile{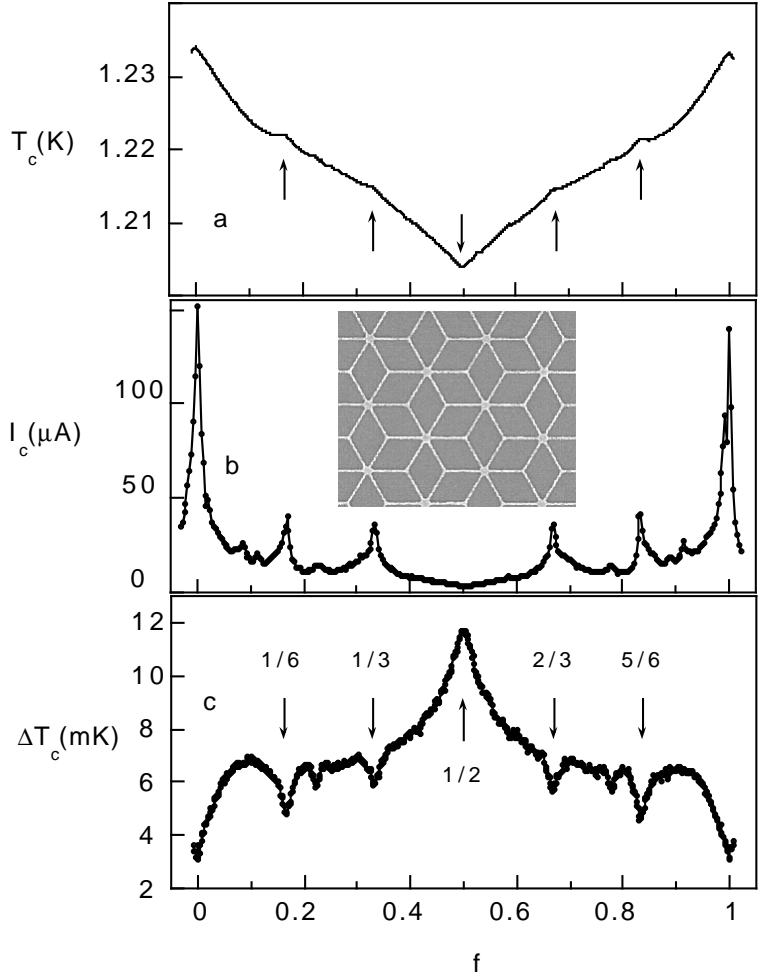}}
\caption{Top : Critical temperature as a function of the 
frustration (frustration $f=1$ corresponds to an applied magnetic field 
of $B = 0.234$ mT). Middle : Critical current at fixed temperature 
$T=1.185$ K. Bottom : width of the superconducting transition. Note the 
behaviour at $f=1/2$ strongly differs from the behaviour at other 
rational frustrations $f=p/q$. Insert shows a micrograph of one part 
of the measured wire network : The aluminum strand are $1\mu$m long 
with a cross section 100 x 100 nm$^{2}$.
} 
\end{figure}

The observed $T_{c}¥(H)$ transition line (Fig1.a) can be understood in its very 
details \cite{Abilio} from the ground state of the Landau level 
spectrum and will not be discussed here.  The critical current curve 
measured at constant temperature, $T=1.185$ K is shown in Fig.1b.  For 
each magnetic field the dynamic resistance characteristics was 
measured $vs$ increasing dc bias current.  The critical current was 
defined as the threshold current for which the dynamic resistance 
exceeds $0.2\%$ of the normal state resistance ($R_{N}$).  The 
width $\Delta T_{c}$
of the superconducting transition (Fig.1c) is defined as the 
difference between $T_{c}$ curves taken from criteria $0.6R_{N}$ and 
$0.1R_{N}$.  The collection of data shown in Fig.1 reveals a quite 
unusual behaviour of the superconducting state of the $T3$ network.  
The usual picture that rational frustrations are characterized by (i) 
an upward cusp in the critical temperature curve, (ii) a sharp peak in 
the critical current and (iii) a dip in the transition width, is 
observed for values $0$, $1$, $1/6$, $1/3$, etc\ldots but 
{\it is not observed} for the simplest rational number $f=1/2$.  
Instead one observes a downward cusp in the critical temperature as in 
the case of an isolated single loop \cite{LP} while the critical 
current shows absolute minimum for this particular value.  The 
suppression of critical current at $f=1/2$ reflects the effect of the 
band structure on the superfluid velocity and provides strong evidence 
of the non dispersive character of the eigenstates.  The large broadening 
of the transition indicates that phase correlation between network 
sites cannot be established, leading to enhanced phase fluctuations.
\\This anomalous behaviour was never pointed out before.  We claim 
that it is related to the localization effect discussed in 
ref\cite{Vidal}.  Further experiments have been carried out recently 
in sample with wire strand lengthes as small as $0.5 \mu$m and will be 
published elsewhere \cite{Serret}.\\  \\ 
The above discussed phenomena have their counterpart on the properties 
of the vortex lattice.  The frustration $f$ is still the only control 
parameter which, here, can be viewed as the filling factor 
for vortices in the plaquettes of the lattice.  We have used the 
Bitter decoration technique to visualize the vortex configuration in 
the $T3$ lattice for $f=1/3$ and $f=1/2$.  As shown recently 
\cite{Bezryadin} the decoration contrast can be significantly 
improved, in the case of superconducting arrays, by using the 
so-called flux compression method in which a thin superconducting 
bottom layer converts the network coreless vortices into well-defined 
Abrikosov vortices.  Fig.2 compares the observed vortex patterns for 
$f=1/3$ and $f=1/2$.  The $1/3$ case represents the reference 
situation where the natural triangular vortex lattice is commensurate 
with the underlying network.  As expected we do observe an ordered state 
(Fig.2 left) which consists of a single domain state with a few 
point defects but without domain walls.  The critical current peak and 
the pinning of the vortex lattice are two related phenomena whose origin 
is the dispersive curve $\epsilon (k)$.  The nature of the 
vortex state at $f=1/3$ in the $T3$ lattice is similar to that of the 
checkerboard commensurate state observed at half filling in a square 
lattice \cite{Runge,Hess}.  In contrast, the $f=1/2$ vortex structure 
does not exhibit any commensurate state (Fig.  2 right).  A detailed 
analysis of the measured vortex correlation functions on an array 
containing several thousands plaquettes \cite{Serret} shows 
\underline {no long range order} under these 
conditions.  Although there is no available theoretical model for the 
nature of the vortex ground state, the absence of a vortex commensurate state 
is likely consistent with the very high degeneracy of the electronic 
states in the tight-binding formulation.  This result may have 
significant relevance on the more general field of frustrated systems.  
A detailed investigation of the ground states of vortices sitting on a 
Kagome lattice which is the dual of the $T3$ lattice is now in 
progress \cite{Butaud}.

\begin{figure}
    \epsfxsize=130mm
		  \centerline{\epsffile{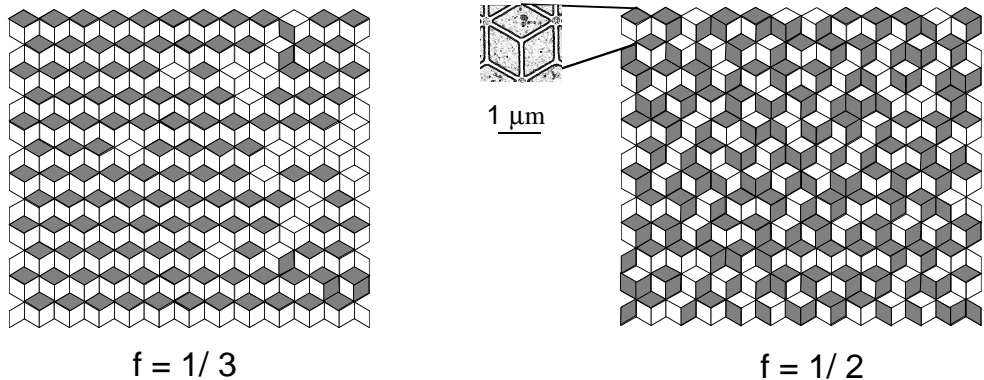}}
		  \caption{The vortex configuration at f=1/3 (left) and 
f=1/2 (right).  Inset : SEM picture showing three cells with one 
decorated vortex clearly visible in the middle of the upper cell. The 
samples for decoration are made of niobium epitaxially grown on a 
sapphire substrate. The elementary strands are $1\mu$m long, 
$100$ nm wide and $130$ nm thick. The $70$ nm thick uniform bottom layer 
allows for unambiguous identification of vortices inside the cells 
(inset) }
\end{figure}

\section{Conclusion}

It is noteworthy that the rationality of the frustration parameter at
$f=1/2$ does not lead to a commensurate state in the $T3$ lattice.  
This uncommon situation is demonstrated
by the decoration experiment.  Actually the properties of the
superconducting state at $f=1/2$ are reminiscent of the case of
irrational frustration \cite{Yu} in a square lattice.  We believe that
the anomalous superconducting properties presented in this paper are
consistent with the localization effect predicted in ref \cite{Vidal}. 
The weakening of phase coherence by destructive interference is
unambiguously observed.  The mapping between the superconducting
properties with the tight-binding problem allows for a formal
connection between the current carrying superconducting states and the
Landau level spectrum.  The observed magnetic field dependence of the
critical current can be accounted for qualitatively \cite{Abilio} by
expressing the supercurrent in terms of the curvature of the
band edge $\epsilon (k)$. However, this mapping is only valid 
in the vicinity of the critical temperature. 
The ordinary superconducting behaviour is recovered at lower temperature
when the superconducting coherence length becomes smaller than the
cell size.  In this regime the large energy barriers prevents 
the vortex motion across the superconducting wires.  A quite
different situation may occur in Josephson junction arrays
\cite{Martinoli} where the superconducting phase fluctuations play a
more important role.  The absence of commensurate states in the $T3$
lattice at $f=1/2$ may lead to interesting dynamics of the vortices at
low temperatures.  Apart from the superconducting systems, arrays of
quantum wires \cite{Mailly} with the $T3$ geometry are also expected to
exhibit interesting properties related to this unique interference
effect.

\ack We acknowledge B. Dou\c{c}ot, R. Mosseri and O. Buisson for 
fruitful discussions.  The \textit{e-beam} lithography and the SEM 
observations were carried out with PLATO organization teams and tools.


\begin{thebibliography}{99}
    
\bibitem{exp} B. Pannetier, J. Chaussy and R. Rammal, J. Phys.  
Lettres {\bf 44}, L853 (1983); B. Pannetier, J. Chaussy, R. Rammal, 
Ph.  Gandit and J.C. Vill\'{e}gier, Phys.  Rev.  Lett.  {\bf 53}, 1845 
(1984).

\bibitem {Gandit} P. Gandit, J. Chaussy, A. Vareille and A. Tissier, 
EuroPhys. Lett. {\bf 3}, 623 (1987).

\bibitem {Buisson} O. Buisson, M. Giroud and B. Pannetier, EuroPhys.  
Lett.  {\bf 12}, 727 (1990).

\bibitem{Vidal} J. Vidal, R. Mosseri and B. Dou\c{c}ot, Phys.  Rev.  
Lett.  {\bf 81}, 5888 (1998).

\bibitem{Hofstadter} D.R. Hofstadter, Phys.  Rev.  B {\bf 14}, 2239 
(1976).

\bibitem{Tinkham} M. Tinkham, {\it Introduction to 
Superconductivity}, MacGraw Hill Inc (1996).

\bibitem {Wang} Y.Y. Wang, R.Rammal and B. Pannetier, J. Phys. {\bf 49}, 
2045 (1988).

\bibitem{Abilio} C.C. Abilio, P. Butaud, Th. Fournier, B. Pannetier, 
J. Vidal, S. Tedesco and B. Dalzotto, Phys.  Rev.  Lett.  {\bf 83}, 
5102 (1999).

\bibitem{Serret} E. Serret et al. to be published.

\bibitem{LP} W.A. Little and R. Parks, Phys.  Rev.  A {\bf 44}, 97 
(1964).

\bibitem{Bezryadin} A. Bezryadin, Y. Ovchinnikov and B. Pannetier, 
Phys. Rev. B {\bf 53}, 8553 (1996).

\bibitem{Runge}  K. Runge and B. Pannetier, EuroPhys. Lett. 
{\bf 24}, 737 (1993).

\bibitem{Hess}H.D. Hallen, R. Seshadri, A.M. Chang, R.E. Miller, L.N. 
Pfeiffer, K.W. West, C.A. Murray and H.F. Hess, Phys. Rev. Lett. {\bf 
71}, 3007 (1993).

\bibitem{Butaud} P. Butaud et al. to be published.

\bibitem{Yu} F. Yu, N. E. Israeloff, A.M. Goldman and R. Bojko, Phys. 
Rev. Lett. {\bf 68}, 2535 (1992).

\bibitem{Martinoli} see for example the recent review by P. Martinoli and Ch. 
Leemann, J. Low Temp. Phys. {\bf 118}, 699 (2000).

\bibitem{Mailly} C. Naud and D. Mailly, Proceedings of the EPS 
conference, Montreux, march 2000.

\end{thebibliography}
\end{document}